\documentclass[12pt,referee]{aastex}
\usepackage{graphicx}
\usepackage{bm}
\usepackage{amsmath}
\usepackage{natbib}

\newcommand{\eqb}{\begin{eqnarray}}
\newcommand{\eqe}{\end{eqnarray}}

\newcommand{\strength}{a}
\newcommand{\rcrit}{r_{\rm cr}}

\slugcomment{}
\shorttitle{Superluminal waves in pulsar winds}
\begin{document}
\title{Superluminal waves in pulsar winds}
\author{Ioanna Arka\footnote{present address:
Institute de Plan\'etologie et d'Astrophysique de Grenoble, UMR 5274,
BP 53 F-38041 Grenoble, France}  and John G. Kirk}
\affil{Max-Planck-Institut f\"ur Kernphysik, Postfach 10~39~80,
69029 Heidelberg, Germany}
\email{ioanna.arka@mpi-hd.mpg.de, john.kirk@mpi-hd.mpg.de}

\begin{abstract}
  The energy lost by a rotation-powered pulsar is carried by a
  relativistic flow containing a mixture of electromagnetic fields and
  particles. In the inner regions, this is thought to be a
  magnetically dominated, cold, electron-positron wind that is well
  described by the MHD equations. However, beyond a critical radius
  $\rcrit$, the same particle, energy and momentum fluxes can be
  transported by a strong, transverse electromagnetic wave with
  superluminal phase speed. We analyze the nonlinear dispersion
  relation of these waves for linear and circular polarization, and
  find the dependence of $\rcrit$ on the mass-loading, magnetization
  and luminosity of the flow, as well as on the net magnetic flux. We
  show that, for most isolated pulsars, the wind lies well outside
$\rcrit$, and speculate that superluminal modes play an
  important role in the dissipation of electromagnetic energy into
  nonthermal particles at the termination shock.
\end{abstract}


\section{Introduction}
Pulsar winds are believed to be launched as relativistic outflows
consisting mainly of electrons and positrons. These winds are
Poynting-flux dominated when they are launched from the pulsar. Outside the 
light-cylinder (located at $\rho=r/r_{\rm LC}=r\Omega/c=1$,
with $\Omega$ the angular velocity of the pulsar) they move radially,
carrying a frozen-in, nearly transverse magnetic field that is modulated by
the pulsar rotation. They can be viewed as nonlinear waves, whose
phase speed equals the radial flow speed (and is, therefore,
subluminal).  The simplest analytic description of this flow is the
\lq\lq striped wind\rq\rq\ \citep{coroniti90}, in
which the magnetic field is purely toroidal, has a magnitude
proportional to $1/r$, and reverses sign across a corrugated current
sheet that separates the two magnetic hemispheres. Numerical solutions
of the force-free, oblique rotator problem \citep{spitkovsky06}, lend
qualitative support to this picture, which has also been used to model
the high-energy emission of pulsars
\citep{petrikirk05,petri11} as well as the orbital modulation of the high
and very high energy emission from gamma-ray binaries that contain pulsars 
\citep{petridubus11}.

In the absence of dissipation, these subluminal waves propagate at
constant speed and remain Poynting dominated, provided their density
is high enough to apply the MHD approximation
\citep{kirkmochol11}. However, basic dynamical considerations indicate
that the winds should not be Poynting dominated after they cross the
termination shock \citep{reesgunn74,begelman98}, and spectral and
morphological modeling of pulsar wind nebulae suggest an even
stricter upper limit on the Poynting dominance. How Poynting flux is
converted to kinetic energy flux is still not clear \citep[for reviews
see][]{gaenslerslane06,kirketal09}.  Reconnection in the current
sheet, as proposed by \citet{coroniti90} and \citet{michel94},
proceeds rather slowly
\citep{lyubarskykirk01,kirkskjaeraasen03,lyubarsky10}. In the case of
an isolated pulsar such as the Crab, it fails to convert a significant
fraction of the electromagnetic flux before the flow reaches the
termination shock, unless the rate of pair injection by the pulsar is
much higher than conventionally assumed \citep{arons11}.

Motivated by this finding, \citet{lyubarsky03} suggested that the
fields are dissipated in the termination shock itself. This is not
permitted at an MHD shock in a strongly magnetized flow, where the
compression ratio remains close to unity. Nevertheless, it is possible
to imagine a viable scenario in which reconnection in the current
sheet of a striped wind is driven or triggered by the relatively weak
compression of the MHD shock
\citep{petrilyubarsky07,Lyubarskyliverts08}.  Recent 2D and 3D~PIC
simulations exhibit such an effect \citep{sironispitkovsky11},
albeit at even higher pair injection rates than those suggested by
\citet{arons11}.
   
An alternative picture is offered by superluminal waves. 
Because the density in the wind falls with increasing radius,
these modes, which are absent in the MHD description, can
propagate outside of a certain critical radius $\rcrit$ 
\citep{usov75,melatosmelrose96}.  If this radius lies inside the
termination shock, the possibility arises that the flow converts from
a subluminal to a superluminal wave either spontaneously at some point
upstream of the shock, or as an integral part of the shock itself. The
latter possibility is an attractive scenario, since these modes
are known to damp much more rapidly than the subluminal striped wind
mode \citep{leelerche78,asseollobetschmidt80}, guaranteeing a relatively 
thin transition region. 

At low amplitudes, transverse, superluminal waves in an
electron-positron plasma have simple properties
\citep[e.g.,][]{iwamoto93}. However, under pulsar conditions, the
waves are strongly nonlinear. Their properties 
have been investigated in many papers, 
\citep[for example][]{kawdawson70,maxperkins71,clemmow74,clemmow77,asseollobetpellat84,melatosmelrose96,skjaeraasenmelatosspitkovsky05}. 
Here we take a new look at the problem, and extend the known results for 
both circularly polarized waves,
and linearly polarized waves with a non-zero (phase-averaged) magnetic
field, using the two-fluid (electron-positron) model. In
particular, we formulate and solve the equations that determine into
which modes a striped wind of given magnetization can convert. This
amounts to solving the electrodynamic equivalent of the jump
conditions across a magnetohydrodynamic shock front. As a result, we
find the cut-off radius of these modes as a function of the striped
wind parameters. We also show that the frequently used 
\lq\lq strong wave limit\rq\rq\  
does not provide an adequate description of these modes.

The paper is set out as follows: in Section \ref{twofluid} we 
review the
basic equations for plane waves. Using the \lq\lq homogeneous\rq\rq\
or \lq\lq H-frame\rq\rq\ formulation introduced 
by \citet{clemmow74,clemmow77}, we compute the fluxes of particles, 
energy and momentum carried as functions of the wave parameters. 
In Section \ref{results}, we use these relations
to solve the jump conditions that must be fulfilled when the wind 
converts from a subluminal to a superluminal mode across a transition 
layer that is narrow compared to its distance from the pulsar. 
Section \ref{discussion} contains a discussion of these
results and their relation to the problem of dissipation at the
termination shock.

\section{Superluminal waves in an electron-positron plasma}
\label{twofluid}

An analytical treatment of non-linear waves in a cold plasma was first
undertaken by \citet{akhiezerpolovin56} who
considered a one-component electron plasma in a background of fixed
ions.  In their approach, all wave quantities depend only on the phase
$\phi$ of the wave. For waves propagating in the $x$-direction, 
$\phi = \omega( t - x/v_{\phi}) $, where $v_{\phi}$ is the phase
velocity and $\omega$ the frequency of the wave. Both subluminal and
superluminal waves ($v_{\phi}<c$ and $v_{\phi}>c$ respectively) are
possible, and the longitudinal and transverse modes are, in general, 
coupled. Pulsars are most likely surrounded by an electron-positron
pair plasma, possibly loaded with a small number of protons, so that a
single-fluid approximation is not valid. Also, the wave
amplitudes near the pulsar are expected to be high enough to drive
even the proton component of a plasma to relativistic velocity, suggesting 
that the rest mass of the fluid particles is unimportant. 
Consequently, rather than considering the general case of an 
electron-ion plasma, attention has been concentrated on 
the much simpler two-fluid electron-positron
plasma. In such a plasma, both linearly and circularly polarized waves
with purely transverse fields can propagate without introducing
fluctuations in the charge-density
\citep{asseokennelpellat75,kennelpellat76}. It is these waves that we
consider in the following.

Far from the pulsar (at $\rho\gg1$), the magnetic field is twisted
into a tightly-wound spiral form, and it is an excellent approximation
to neglect the phase-averaged radial field component compared to the
toroidal one. Since pulsar winds are radial, this means that 
the phase-averaged component of the magnetic
field is transverse. Furthermore, in this region, 
the radius of curvature of the
wavefront is much larger than one wavelength, so that the waves
can be approximated as locally plane. 

In the following, we identify the (radial)
propagation direction of the wave as the $x$-axis
and the toroidal direction as the $z$-axis, so that $E_x=0$ and
$B_x=0$. The equations to be solved are the continuity equations
and the equations of motion for each of the two particle species,
along with Maxwell's equations. These are given in
Appendix~\ref{2fluideqns}, where it is shown that, from Maxwell's
equations and making use of the assumption of purely transverse waves, the 
two species
have the same density: $\gamma_+n_+ = \gamma_-n_-$ (a plus (minus) subscript 
denotes positrons (electrons)). Furthermore, the Lorentz factors are equal:
$\gamma_+ = \gamma_-$ and the other components of the dimensionless four-velocity of the
two species are related by:
\begin{eqnarray}
\nonumber
u_{x+} &=& u_{x-} \quad u_{y+} \,=\, -u_{y-} \quad u_{z+} \,=\, -u_{z-}
\enspace.
\end{eqnarray}
To keep the notation simple, we henceforth 
drop the plus and minus subscripts and 
express everything in terms of the positron fluid quantities.

For superluminal waves, \citet{clemmow74,clemmow77} showed the problem
is significantly simplified if one performs the calculations in a
frame of reference that propagates with a speed $c^2/v_{\phi}$ with
respect to the lab.\ frame.  In this frame, which we call the
H-frame, the wave quantities do not depend
on space but only on time, and the phase of the wave is just $\phi =
\omega t$.  An immediate consequence of Faraday's equation is that in
this frame the magnetic field, in addition to being homogeneous, is
also constant. In the following, unprimed symbols denote quantities in
the H-frame; we use primed symbols for quantities in the lab.\ frame.

\subsection{Circular polarization}
The circularly polarized wave is described in the H-frame by three
(phase-independent) parameters: the (constant) proper number density of each species
$n=n_0$, the
component of the four-velocity in the direction of propagation
$u_x=u_{x0}$, and the magnitude of the four-velocity component perpendicular 
to the propagation direction $\sqrt{u_y^2+u_z^2}$, which, for this wave mode, 
equals the dimensionless 
{\em strength parameter} denoted by $a$ and defined in terms of the 
electric field amplitude $E_0$ as 
\begin{eqnarray}
a&=&e\left|E_0\right|/\left(mc\omega\right)
\enspace.
\label{strengthdef}
\end{eqnarray}
Consequently, the Lorentz factor of the fluids
is also phase-independent: $\gamma=\gamma_0=\sqrt{1+u_{x0}^2+a^2}$.
The frequency $\omega$ 
coincides with the (proper) plasma frequency, $\omega_{\rm p}$: 
\begin{eqnarray}
\omega&=&\omega_{\rm p}\,=\, \sqrt{\frac{8\pi n_0 e^2}{m}} 
\enspace.
\label{plasmafreq}
\end{eqnarray}

In terms of these quantities, the particle flux density $J$ and the $(0,0)$,
$(0,x)$ and $(x,x)$ components of the stress energy tensor are given
by (see Appendix~\ref{lorentzappendix}, equations~\ref{appT1}-\ref{appT3} and 
\ref{appTem1}-\ref{appTem3}):
\begin{eqnarray}
J       &=& n_0\gamma_0 c    \left(\frac{2 u_{x0}}{\gamma_0}\right)\\
T^{00}   &=& n_0\gamma_0 mc^2 \left(2\gamma_0 + \frac{a^2}{\gamma_0}\right)\\
T^{01}   &=& n_0\gamma_0 mc^2 \left(2u_{x0}\right)\\
T^{11}   &=& n_0\gamma_0 mc^2 \left(\frac{2u_{x0}^2}{\gamma_0} + 
\frac{a^2}{\gamma_0}\right)
\end{eqnarray}
and are also phase-independent.

\subsection{Linear polarization}

The linearly polarized mode requires an additional parameter to fix
the amplitude of the phase-averaged magnetic field.  In the H-frame,
the four parameters are: the proper number density $n_0$ and the
component $u_{x0}$ of the four velocity in the direction of
propagation {\em both taken at phase zero}, (the point in the wave
where the magnitude of the electric field is maximal, and $u_z = u_y =0$),
the nonlinearity parameter $q$, used by \citet{kennelpellat76},
and the ratio $\lambda$ of the (phase-independent) magnetic field to
the electric field $E_0$ at phase zero. Physically $q$ is double the ratio 
at phase zero of
the energy density in the fluids to that in the electric field, as measured
in the H-frame:
\begin{eqnarray}
q&=&\frac{32\pi n_0\gamma_0^2mc^2}{E_0^2}
\enspace.
\label{qdefinition}
\end{eqnarray}
Defining the normalized field variable $y=E/E_0$, one finds:
\begin{eqnarray}
u_x&=&u_{x0}+4\gamma_0(1-y)\lambda/q
\label{uxeqlin}
\\
\gamma&=&\gamma_0+2\gamma_0\left(1-y^2\right)/q
\label{gammaeqlin}\\
u_y &=& 0 \\
\left| u_z\right|&=&\sqrt{\gamma^2-u_x^2-1}
\label{uzeqlin}
\nonumber\\
&=&2\gamma_0\sqrt{N(y)}/q
\end{eqnarray}
where $N(y)$ is a fourth-order polynomial, as follows from 
(\ref{uxeqlin}) and (\ref{gammaeqlin}), 
and $\gamma_0=\sqrt{1+u_{x0}^2}$. Periodic
wave solutions exist provided $N(y)$ has four distinct real roots
$y_{1\dots4}$, in which case $y$ oscillates between the values 
$y_2$ and $y_3$ (assuming the ordering $y_1<y_2<y_3<y_4$).
Because of the normalization of $y$, one has $y_3=1$ and 
$\left|y_2\right|\le1$.
The strength parameter of the wave, as defined in (\ref{strengthdef}) is
\begin{eqnarray}
a&=&\frac{2}{\pi}
\int_{y_2}^1\,dy\,\gamma/\sqrt{N(y)}
\label{strengthlin}
\end{eqnarray}
and the wave frequency is given by
\begin{eqnarray}
\omega&=&\frac{2\gamma_0}{a\sqrt{q}}\omega_{\rm p}
\enspace.
\label{linpolfrequency}
\end{eqnarray}
The phase-averaged fluxes are:
\begin{eqnarray}
\left<J\right>       &=& n_0\gamma_0 c    \left<\frac{2 u_x}{\gamma}\right>
\label{javerage}\\
\left<T^{00}\right>   &=& n_0\gamma_0 mc^2 \left<2\gamma + \frac{4\gamma_0\left(y^2+\lambda^2\right)
}{q}\right>
\label{eaverage}\\
\left<T^{01}\right>   &=& n_0\gamma_0 mc^2 \left<2u_x + \frac{8\gamma_0y
\lambda}{
q}\right>
\label{faverage}\\
\left<T^{11}\right>   &=& n_0\gamma_0 mc^2 \left<\frac{2u_x^2}{\gamma} + 
\frac{4\gamma_0\left(y^2+\lambda^2\right)}{q}\right>
\label{qaverage}
\end{eqnarray}
and the phase-averaged electric field is
\begin{eqnarray}
\frac{\left< E\right>^2}{4\pi}&=&n_0\gamma_0mc^2
\left(\frac{4\gamma_0 \left<y\right>^2}{q}\right)
\enspace.
\label{magneticflux}
\end{eqnarray}
Amp\`ere's law enables these averages to be expressed as integrals
over the normalized electric field $y$:
\begin{eqnarray}
\left<A(y)\right>&=& \frac{\int_{y_2}^1\,dy\,A(y)\gamma/\sqrt{N(y)}}
{ \int_{y_2}^1\,dy\,\gamma/\sqrt{N(y)}}
\end{eqnarray}
(see Appendix~\ref{2fluideqns}). 

If the phase-averaged magnetic field vanishes ($\lambda=0$) one
finds $y_2=-1$, and the integrals in
(\ref{linpolfrequency})--(\ref{magneticflux}) 
can be expressed in closed form in
terms of elliptic integrals. For $\lambda\ne 0$, closed forms can be
found to lowest order in an expansion in the small parameter $q$
\citep[see][]{kennelpellat76}. However, these are not adequate to
describe the pulsar case, as we show in the following section. In the
general case, one must resort to numerical integration, noting that the
integrands have integrable singularities at each of the limits.

\subsection{Conserved quantities}

In order to identify those parts of the wind of a given pulsar in
which superluminal waves can propagate, one has to associate
the wave properties with wind quantities inferred from the
observational data. A pulsar wind transports particles, energy and
magnetic flux, and exerts a ram-pressure on its surroundings. 
Across a stationary shock front the quantities
$J'$, $T'^{01}$ and $T'^{11}$ are conserved. Also, from 
Faraday's law, the transverse electric field component $E'$ is conserved. 
Across a thin transition front where one wave-mode converts into another, 
and which is stationary on the oscillation timescale, the 
{\em phase-averaged} 
values of these quantities are conserved. 
 
However, it is the relative magnitudes of these quantities,
rather than their absolute values, that determine the physics
of wave propagation.
We therefore introduce three dimensionless quantities:
\begin{eqnarray}
\mu = \frac{\left< T'^{01} \right>}{mc \left< J' \right>} \label{mugeneral}\\
\nu = \frac{\left< T'^{11} \right>}{mc \langle J' \rangle} \label{nugeneral}\\
\eta =  \frac{\langle E' \rangle^2/4\pi}{mc \langle J' \rangle} 
\label{etageneral}
\end{eqnarray}
which are conserved across a transition layer and are 
independent of the density parameter $n_0$.  
$\mu$ is the mass-loading parameter introduced by \citet{michel69}, which 
corresponds to the Lorentz factor each particle would need if they were 
to carry the entire energy flux. 

For the superluminal wave modes discussed above, these quantities can
be computed by Lorentz transforming their values in the H-frame into
the pulsar, or lab.\ frame. A boost of (dimensionless) speed 
$\beta_>=c/v_{\phi}$ in the negative $x$ direction 
is required, for which explicit expressions are
given in Appendix~\ref{lorentzappendix}. 
(We use the notation $\beta_>$ and $\Gamma_>$ for the speed and 
associated Lorentz factor of the H-frame of the 
superluminal wave, and 
$\beta_<$ and $\Gamma_<$ for the phase (or bulk) speed and associated 
Lorentz factor of the 
subluminal wave, both as
seen from the lab.\ (pulsar) frame.) 
Applying these boosts, one
obtains, for linearly polarized modes, expressions for $\mu$,
$\nu$ and $\eta$, as functions of the
four input parameters $u_{x0}$, $q$, $\lambda$ and $\beta_>$
that must be evaluated by numerical integration. The case
of circular polarization, where $\eta=0$, is much simpler, since
$\mu$ and $\nu$ are given in closed form as functions
of $u_{x0}$, $a$ and $\beta_>$. 

For the cold subluminal wave modes, such as the striped wind
(in the absence of dissipation), $\mu$ and $\nu$ are 
independent of radius in the inner parts of the wind where 
a single-fluid MHD description is adequate
\citep{kirkmochol11}. Their values 
can be estimated from standard pulsar models
\citep[e.g.,][]{lyubarskykirk01}. However, these two quantities are
almost equal in highly relativistic winds, and it proves more convenient 
to use, instead of $\nu$, the magnetization parameter $\sigma$, defined 
 as the ratio of magnetic to particle energy flux according to:
\begin{eqnarray}
\mu &=& \Gamma_< (1+\sigma) \label{mumhd} \\
\nu &=& \frac{\Gamma_<^2 (1+\sigma) -  
(1+\sigma/2)}{\sqrt{\Gamma_<^2-1}} \label{numhd}
\end{eqnarray}
In the striped wind, the third
parameter, $\eta$, depends on the phase-averaged magnetic field value
\begin{eqnarray}
b&=&\left<B'\right>/\left<B'^2\right>^{1/2}
\label{bdefinition}
\enspace,
\end{eqnarray}
via the expression:
\begin{eqnarray}
\eta&=& b^2\sigma \beta_<\Gamma_< \label{etamhd}
\enspace.
\end{eqnarray}
Without loss of generality, we may choose $\left<B'\right>\ge 0$, so that
$0 \le b \le 1$.
Both $b$ and $\eta$ are
functions of latitude in the striped pulsar wind, since the spacing in
phase of the current sheets varies. At latitudes greater than
the inclination angle between the magnetic and rotation axes, 
conventionally denoted by $\alpha$, the sheets vanish 
(zero spacing), whereas on the equator, they are equally spaced 
(separated in phase by $\pi$).

Thus, for given $\mu$, $\sigma$ and $b$, the jump conditions
can be solved for three of the 
four input parameters $u_{x0}$, $q$, $b$ and $\beta_>$, 
provided the fourth is specified. The wave frequency, normalized to 
$\omega_{\rm p}$, can then be constructed from (\ref{linpolfrequency}).

\section{Results}
\label{results}

\subsection{Circular polarization}

\begin{figure}
\includegraphics[scale=1.35]{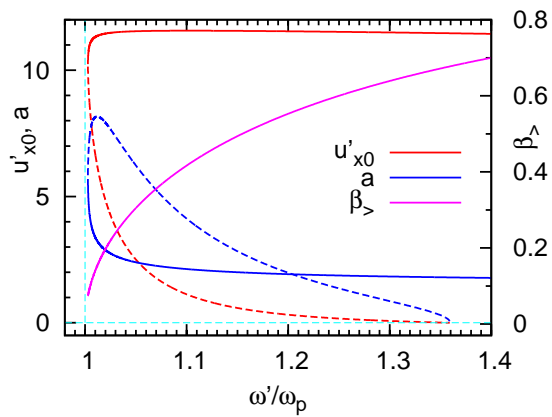}
\caption{The circularly polarized modes that satisfy the jump
  conditions for $\mu=12$, $\sigma=3$ as a function of lab.\ frame
  frequency normalized to the proper plasma frequency. No solutions
  exist below a cut-off frequency that is close to 
$\omega'/\omega_{\rm
    p}=1$. Two solutions, one with $u_{x0}\approx\mu$ and one with small
  $u_{x0}$ (denoted by solid and dashed lines, respectively) exist
  above the cut-off, up to
  $\omega'/\omega_{\rm p}\approx1.36$, where
  $a\rightarrow0$ for the low-$u_{x0}$ mode. The modes have the same
  refractive index for given $\omega'/\omega_{\rm p}$, 
which equals
  the speed $\beta_>$ of the H-frame seen from the lab.\ frame.}
\label{circvsomega}
\end{figure}
For circular polarization, the jump conditions can be solved
analytically \citep{kirk10}. An example is plotted in
Fig.~\ref{circvsomega}, which shows the refractive index
$ck'/\omega'=\beta_>$, the strength parameter $a$ and the component of 
the particle speed in the propagation direction, $u_{x0}$, 
as a function of lab.\ frame frequency
$\omega'$ normalized to the (proper) plasma frequency $\omega_{\rm p}$. 
In this example, we have chosen moderate values of the energy
and momentum fluxes in the lab.\ frame: $\mu=12$,
$\sigma=3$ (and, therefore, $\Gamma_<=3$), 
in order to display the structure of the
solutions near the frequency cut-off. Since the wavenumber $k$ vanishes in the 
H-frame, the refractive index is trivially
obtained from the Lorentz transformation of the wave number and
frequency: $\omega'=\Gamma_>\omega$ and $k'=\beta_> \Gamma_>\omega/c$, 
(this also applies, of course, to the case of linear polarization)
and the dispersion relation $\omega=\omega_{\rm p}$ implies 
$\omega'/\omega_{\rm p}=\Gamma_>$.
However, the corresponding strength
parameter $a$ is found only after solving the full set of
equations. As can be seen from this figure, there are no physical
solutions for $\omega'<\omega_0$, two solutions for
$\omega_0<\omega'<\omega_1$ and only one physical solution for
$\omega'>\omega_1$.  (The full expressions for $\omega_{0,1}$ are
cumbersome, but for $\mu\gtrsim \sigma^{3/2}\gg1$
(corresponding to a mildly supermagnetosonic outflow), one finds
$\omega_0\approx\omega_{\rm
  p}\left[1+\left(4\mu^2-\sigma^3\right)\sigma /\left(32\mu^4\right)
+\textrm{O}\left(\mu^{-5/3}\right)\right]$
  and $\omega_1\approx \omega_{\rm p}
\left(\mu/8\right)^{1/4}\left[1+\textrm{O}\left(\mu^{-1/3}\right)\right]$, 
\citep[cf.][]{kirk10}.)  
Thus, the imposition of a finite
  energy flux per particle prohibits wave propagation at frequencies
  close to the cut-off $\omega'=\omega_{\rm p}$ found from linear
  theory. At intermediate frequencies, waves of two different
  amplitudes, and different values of $u_{x0}$ are available to
  transport the required fluxes.  One of these disappears
  ($a\rightarrow0$) at finite $\omega'/\omega_{\rm p}$, leaving only a
  single solution of the jump conditions at high frequency.

In the case of a pulsar wind, the wave frequency is dictated by the
rotation of the neutron star, but the plasma density decreases with
increasing distance from the star. Thus, rather than solving the jump
conditions as a function of frequency, more insight can be gained by
solving as a function of radius.  The connection between flux density
and radius is found in terms of the luminosity per unit solid angle of
the (radial) wind, $dL/d\Omega_{\rm s}$:
\begin{eqnarray}
\left<F'\right>&=& \frac{1}{r^2}\frac{dL}{d\Omega_{\rm s}}
\enspace,
\end{eqnarray}
which can be expressed in dimensionless form in terms of the quantity 
\begin{eqnarray}
a_{\rm L}&=&\sqrt{\frac{4\pi e^2}{m^2c^5}\frac{dL}{d\Omega_{\rm s}}}
\enspace.
\end{eqnarray}
Physically, $a_{\rm L}$ is the strength parameter of the circularly
polarized vacuum wave that would be needed to carry the entire pulsar
luminosity at the surface $\rho=1$. For a spherically symmetric wind
$a_{\rm L}=3.4\times10^{10} L_{38}^{1/2}$, where
$L_{38}=L/\left(10^{38}\,\textrm{erg\,s}^{-1}\right)$.
This quantity is related to the 
{\em multiplicity} parameter $\kappa$, defined as the Goldreich-Julian
charge density at the light cylinder divided by $e$ 
\citep[see, for example][]{lyubarskykirk01,kirkmochol11} by 
\begin{eqnarray}
\kappa&=& a_{\rm L}/\left(4\mu\right)
\label{kappadefinition}
\end{eqnarray}

Writing the energy flux density as $\mu\left<J'\right>$
leads to the expression
\begin{eqnarray}
\left<J'\right>&=&
\frac{2 n_0c\Gamma_>^2\omega^2 a_{\rm L}^2}{\rho^2\omega_{\rm p}^2\mu}
\enspace.
\end{eqnarray}

Applying this to the case of circular polarization, where
$\left<J'\right>=2n_0u'_{x0}$ and $\omega=\omega_{\rm p}$, one finds
$u'_{x0}=\Gamma_>^2a_{\rm L}^2/\left(\mu \rho^2\right)$. Then,
the inequalities $u'_{x0}<\gamma'\le \mu$ and $\Gamma_>>1$ imply that
circularly polarized waves propagate only when $\rho>a_{\rm  L}/\mu$. We 
therefore introduce the dimensionless scaled radius
\begin{eqnarray}
R&=&\rho\mu/a_{\rm L}
\nonumber\\
&=&
\rho\kappa/4
\end{eqnarray} 
which can be constructed after solving 
the jump conditions from the expression
\begin{eqnarray}
R&=& \frac{\Gamma_>\omega}{\omega_{\rm p}}
\sqrt{\frac{2\mu}{\gamma_0\left(\left<J'\right>/n_0\gamma_0\right)}}
\end{eqnarray}
for both circularly and linearly polarized modes.

\begin{figure}
\includegraphics[scale=1.2]{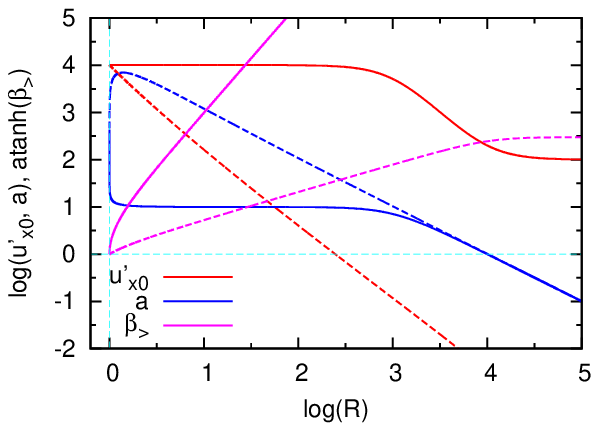}
\caption{The properties of circularly polarized waves 
for $\mu=10100$, $\sigma=100$, as a function of 
dimensionless radius in the pulsar wind. 
The inverse hyperbolic tangent of the refractive index 
$ck'/\omega'=\beta_>$ is 
plotted, in order to bring out both the 
nonrelativistic regime, $\beta_>\ll 1$, 
near $R=1$ and the relativistic regime,
$\Gamma_>\gg1$, at $R\gg1$.}
\label{circulardispersion}
\end{figure}

Figure \ref{circulardispersion} shows the properties of circularly
polarized waves as functions of $R$ for parameters appropriate
to pulsar winds: $\sigma=100$ and $\mu=10100$ (and, therefore,
$\Gamma_<=100$).
Near the cut-off radius, which lies close to $R=1$, 
the mode properties are similar to those
seen in Fig.~\ref{circvsomega}, except that the refractive index,
now plotted as as a function of radius, is no longer degenerate. The detailed
properties of these solutions have been discussed in \cite{kirk10}. However, 
note that these curves do not describe the radial evolution of a wave packet,
as erroneously suggested in that paper and assumed
by \citet{asseollobetpellat84}
(who considered linear polarization), 
but simply specify the wave modes into 
which a wind of given $\mu$ and $\sigma$ can 
convert at a given radius. The difference arises because 
the radial momentum flux density, $\nu$, is not a 
conserved quantity in the radial evolution of the superluminal modes
\citep[see also][]{kirkmochol11a}.

\subsection{Linear polarization}

\begin{figure}
\includegraphics[scale=1.2]{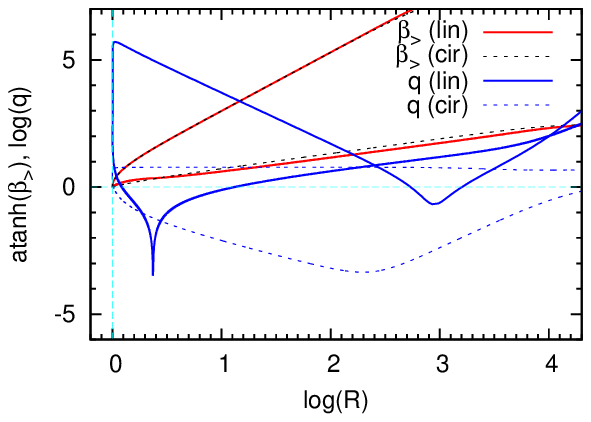}
\caption{A comparison of the refractive indices for circularly and linearly
polarized modes for $\mu=10100$, $\sigma=100$. The nonlinearity 
parameter $q$, defined in (\ref{qdefinition}), is also plotted
for each mode.}
\label{comparisonlincirc}
\end{figure}

When the phase-averaged magnetic field vanishes, the properties of linearly
polarized modes are similar to those of circularly polarized modes. This
is illustrated in Fig.~\ref{comparisonlincirc}, where one sees that
the refractive indices are practically identical. 

This figure also
shows the nonlinearity parameter $q$.  For linear polarization, $q$ is
defined in terms of quantities measured at phase zero. This point in
the wave is, however, special. It corresponds to the turning points of
the \lq\lq saw-tooth\lq\lq\ waveform \citep{maxperkins71}, where the
fluid velocity either vanishes or lies precisely in the propagation
direction. These points do not play a large role in determining the
average properties of the wave.  As a result, $q$ varies
substantially, especially close to points where $u_{x0}$ changes sign
($\log(R)\approx 0.4$ in the figure). One consequence is that an
expansion of the fluxes using $q$ as a small parameter is inadequate
over most of the range relevant for pulsars, despite the fact that the
waves are highly nonlinear
(see the discussion in \S\ref{discussion}). 
In the case of circular polarization, the
quantities that enter into the definition of $q$
(Eq.~\ref{qdefinition}) are phase-independent. The value of $q$ is
a useful characterization of the mode in this case, and  
Fig.~\ref{comparisonlincirc} clearly illustrates that one of the two
solutions for given $R$ is a relatively weak wave, with $q\sim\,$few, whereas
the other is stronger, with $q\sim10^{-2}$. 

\begin{figure}
\includegraphics[scale=1]{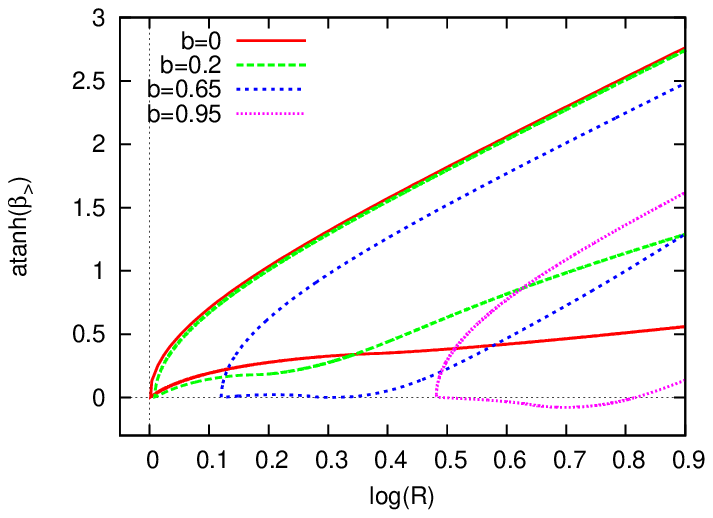}
\caption{The inverse hyperbolic tangent of the refractive index $\beta_>$ for
linearly polarized waves that correspond to an incoming striped wind
with $\mu=10100$, $\sigma=100$, and 
four different values of 
the phase-averaged magnetic field $b$, as
defined in Eq.~(\ref{bdefinition})}
\label{lineardispersion}
\end{figure}

The properties of linearly polarized waves which carry a 
non-zero phase-averaged magnetic field in the toroidal direction 
are shown in Fig.~\ref{lineardispersion}.
The refractive index is plotted for 
superluminal waves that correspond to $\mu=10100$, $\sigma=100$, and 
four 
different values of the parameter $b$, that describes the 
relative amount of phase-averaged magnetic flux carried in the subluminal
wave (see Eq.~\ref{bdefinition}). 
It can be seen that the
cutoff moves to greater $R$ as $b$ rises. 
In the striped wind model, this 
corresponds to moving away from
the equator, where $b=0$.  

Although the refractive indices for non-zero $b$ have similar shape to those
for $b=0$, there is one potentially important difference: for $b>0.65$ 
the lower branch cuts across the $\beta_>=0$ axis, and continues
towards solutions with negative
phase speed. Zero refractive index indicates that the H-frame coincides 
with the lab.\ frame. These solutions are standing waves in the pulsar wind, 
with constant magnetic field and wavelength
$\gg r_{\rm LC}$. Negative refractive index indicates propagation
towards the pulsar, in the sense that the velocity of the H-frame 
is inward propagating. Nevertheless 
these modes still carry the particle, energy, momentum and magnetic 
fluxes outwards towards the nebula.

\begin{figure}
\includegraphics[scale=1]{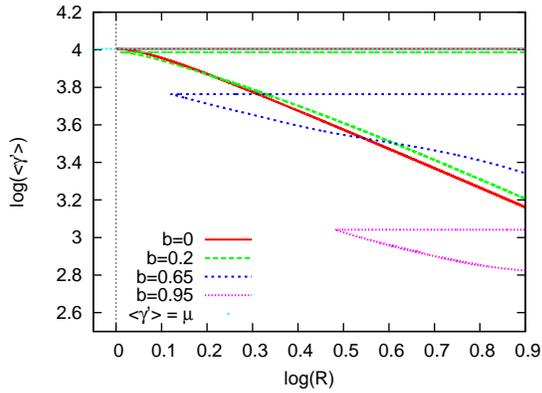}
\caption{The mean Lorentz factor of the outflow particles $\langle
  \gamma' \rangle$ measured in the lab.\ frame, 
for the linearly polarized superluminal wave
  solutions. Particles in a cold MHD outflow 
with the same energy and momentum flux 
have a Lorentz factor of $100$ in this example
($\mu=10100$, $\sigma=100$).
   For the smaller
  $b$ values, $\left<\gamma'\right>\sim \mu$, implying
  a substantial fraction of the field energy has been transferred 
to the particles.}
\label{gammalin}
\end{figure}

Conversion of a subluminal wave into a superluminal wave 
is accompanied by a substantial transfer of energy from the fields to the 
particles. This is quantified in Fig.~\ref{gammalin}, where the 
phase-averaged
Lorentz factor $\langle \gamma'\rangle$ seen in the lab.\ frame
is plotted as a function
of $R$, for the same values of $\mu$, $\sigma$, and 
$b$ as in Fig.~\ref{lineardispersion}. 
For the smaller $b$ values, $b=0$ and $b=0.2$,
the upper branch of the solution has $\langle \gamma'
\rangle \approx \mu$, which implies that almost all 
of the field energy is converted to particles in the 
equatorial regions of the wind. At higher
latitudes, only a part of the Poynting flux can be converted, since the
phase-averaged magnetic flux cannot be dissipated. 
However, even at $b=0.95$, the
upper branch of the dispersion relation carries 
particles with a mean Lorentz factor
one order of magnitude larger than those of the incoming,  
striped wind solution.

\subsection{Latitude dependence}

According to the above figures, superluminal waves cannot propagate in
the inner parts of a pulsar wind. The radius at which propagation
becomes possible is close to $R=1$ for vanishing phase-averaged magnetic field,
and rises with this quantity. However, in the striped wind model,
the luminosity of the wind, and, hence, the parameter $a_{\rm L}$ is
a function of latitude. 
In particular, the energy flux carried by the fields is proportional to
$\sin^2\theta$, 
where $\theta$ is the colatitude, (measured from the rotation axis).   
The flux carried by the particles, on the other hand, is expected to 
be much smaller, but its angular dependence is uncertain. Assuming, for 
simplicity, that the ratio $\sigma$ of these fluxes is independent of 
colatitude, one has
\begin{eqnarray}
a_{\rm L}&=&a_{\rm L, eq}\sin\theta
\end{eqnarray}
$a_{\rm L, eq}=4.2\times10^{10} L_{38}^{1/2}$, is the value of $a_{\rm L}$ in 
the equatorial plane. 

The dependence of the average magnetic field carried by the striped wind
on latitude depends on the inclination angle $\alpha$ between the magnetic and 
rotation axes. Two current sheets are
present in the striped wind, located at phases
\begin{eqnarray}
\phi&=&\pm \phi_{\rm sheet}+2n\pi
\nonumber\\
&=&\pm
\textrm{arccos}\left(\textrm{cot}\alpha\,\textrm{cot}\theta\right)
+2n\pi\enspace,
\end{eqnarray}
where $n$ is an integer \citep{kirketal02}. 
At $\theta=\textrm{atan}(\textrm{cot}\alpha)$, the sheets
merge, and vanish at higher latitudes (smaller colatitudes $\theta$). 
At these sheets, the magnetic field reverses direction, so that
\begin{eqnarray}
b(\theta,\alpha)&=&\left(-\int_{0}^{\phi_{\rm sheet}}\,d\phi 
+\int_{\phi_{\rm sheet}}^\pi\,d\phi\right)/\pi
\nonumber\\
&=&1-2\textrm{arccos}\left(\textrm{cot}\alpha\,\textrm{cot}\theta\right)/\pi
\end{eqnarray}
At those latitudes where there are no sheets, the striped wind model
predicts an azimuthally symmetric magnetic field, i.e., no wave.

\begin{figure}
\includegraphics[scale=01]{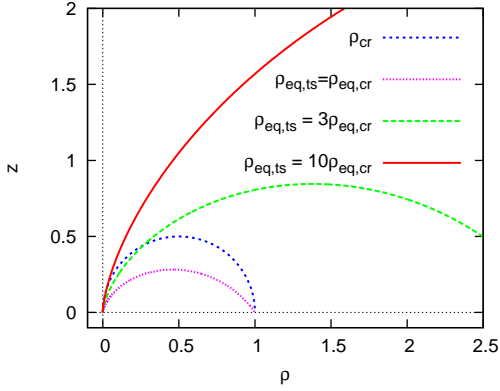}
\caption{ The surface $\rho_{\rm cr}$ outside of which superluminal
  waves propagate is shown for a perpendicular rotator ($\alpha =
  \pi/2$).  The pulsar is at the origin and $z$ is the rotation
  axis. For comparison, the termination shock is also plotted for
  three different external pressures, such that its equatorial radius,
  normalized to that of $\rho_{\rm cr}$, is 1, 3 and 10.}
\label{boundary}
\end{figure}

\begin{figure}
\includegraphics[scale=1]{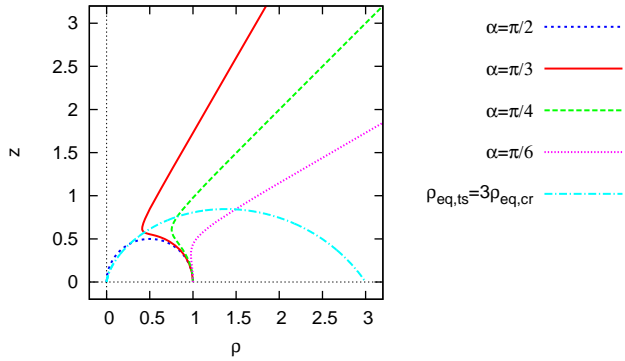}
\caption{
The critical surface $\rho_{\rm cr}$ for various 
inclination angles $\alpha$ between the magnetic
and rotation axes, superimposed on the termination shock surface whose 
equatorial radius is 3 times that of the critical surfaces.}
\label{boundaries}
\end{figure}

From curves similar to those plotted in Fig.~\ref{lineardispersion} 
it is possible, for each value of $b$,  
to identify the critical radius inside which the mode
cannot propagate. This can then be interpolated to give 
a function $R_{\rm cr}(b)$. Together with the 
definition $R_{\rm cr}=\rho_{\rm cr}\mu/a_{\rm L}$, this leads to 
\begin{eqnarray}
\frac{\mu \rho_{\rm cr}}{a_{\rm L, eq}}&=&\sin\theta R_{\rm cr}
\left[b(\theta,\alpha)\right]
\enspace.
\end{eqnarray}
This surface is plotted in Fig.~\ref{boundary}, for the case of the 
perpendicular rotator (magnetic inclination angle $\alpha = \pi/2$), and, in 
Fig.~\ref{boundaries}, for various values of $\alpha$. 

The location of the termination shock is determined by the balance between
the momentum flux carried by the wind and the external pressure. If we 
make the reasonable assumption that the latter is independent of latitude, 
the approximate location $\rho_{\rm ts}(\theta)$ of the shock is given 
for $\sigma\gg1$ by \citep[][Eq.~9]{lyubarsky02}:
\begin{eqnarray}
\left(\frac{d \rho_{\rm ts}}{d\theta}\right)^2+\rho_{\rm ts}^2&=&\rho_0^2\sin^2\theta
\end{eqnarray}
where $\rho_0$ is a constant. 
The normalization of the radius of the termination 
shock surface depends on the 
external pressure and has to be calculated for each individual 
object. In Fig.~\ref{boundary} three examples are shown, 
labeled by the ratio of the equatorial radius of the 
shock to that of the critical surface, $\rho_{\rm eq,ts}/\rho_{
\rm eq,cr}$. It can be seen that if this ratio is significantly 
larger than unity, the cut-off surface falls almost entirely 
within the termination shock, except for a very small region close to the 
pole, where both surfaces have a cusp. A qualitatively similar 
conclusion applies also for different magnetic inclination 
angles, as shown in Fig.~\ref{boundaries}: for  $\rho_{\rm eq,ts}/\rho_{\rm eq,cr} 
\gtrsim$ a few, almost the whole critical surface falls within the 
termination shock radius, apart from a small region at latitudes close
to the magnetic inclination angle $\alpha$.
The plots have been calculated 
using $\mu =10100$ and $\sigma = 100$, and show  one 
quadrant of the poloidal plane. The 3D surface
is generated by reflection in the $\rho$ axis and rotation about the 
$z$-axis. 

When the cut-off surface 
falls inside the termination shock, 
superluminal waves can propagate in the wind. It 
is therefore possible, if the ratio  $\rho_{\rm eq,ts}/\rho_{
\rm eq,cr}>1$, that different modes coexist in the same 
wind but at different latitudes. The MHD 
description is more appropriate for the 
higher latitudes (close to $\alpha$), and the 
superluminal wave solutions exist 
closer to the equatorial plane of the wind. 
If  $r_{\rm eq,ts}/r_{\rm eq,cr} < 1$,
no superluminal modes can be supported by the wind at any latitude. In the
extreme case where $r_{\rm eq,ts}/r_{\rm eq,cr} \gg 1$, practically the whole
wind can support these waves.

The above calculations have been conducted for $\mu\approx 10^4$. In this case
the critical surface in the equatorial plane is located at 
$\rho_{\rm cr}\sim 10^7$ for the Crab pulsar, whereas the termination shock
is located at roughly $\rho_{\rm ts} \sim 10^9$. 
For the Vela pulsar, the critical radius lies at $\rho_{\rm cr}\sim 10^6$,
which is also well within the termination shock, located for this pulsar
at $\rho_{\rm ts}\sim 10^8$ 
\citep[]{kargaltsevpavlov08}. 
This value of $\mu$, corresponds to pair multiplicities of 
$\kappa=2.5\times10^6$ and $\kappa=2.5\times10^5$ for the Crab and Vela, 
respectively, much higher than conventional estimates
\citep[][]{dejager07}, which would place the critical surface even closer
to the pulsar. Thus, at least for 
young, energetic pulsars, 
$\rho_{\rm eq,ts}/\rho_{\rm eq,cr} \gg 1$ and superluminal waves can 
propagate in essentially the entire equatorial wind. 

However, the situation is different for pulsars which are members of binary 
systems, since the wind of the companion star can provide an obstacle 
capable of sustaining the termination shock relatively close to the pulsar.
One example is the system PSR~B1259-63, containing a 48ms pulsar in an 
eccentric orbit about a B2e star. At 
periastron, the separation of the stars is $\rho\sim 4.3\times 10^4$, while at 
apastron it becomes $\rho\sim 6\times 10^5$. 
Very high-energy (TeV) gamma-rays were predicted to emerge from the 
termination shock of this pulsar \citep{kirketal99}, and subsequently 
were detected 
\citep{aharonianetal05,aharonianetal09,abdoetal11} at binary phases 
close to periastron.
Assuming $\mu\approx10^4$ in
this young pulsar, the critical radius at the equator is 
$\rho_{\rm cr}\sim 2.4\times 10^5$, which is greater than periastron 
but smaller than 
apastron separation. Therefore, if the termination shock created by 
the interaction of the stellar wind 
with the pulsar wind lies close to the B2e star, superluminal modes may 
play a role in determining its properties near apastron, but not 
near periastron. It would, therefore, 
be interesting to search for an observable diagnostic
of this transition, such as an orbital 
modulation of the radiation from the shock front.
  
\section{Discussion}
\label{discussion}

Nonlinear superluminal waves in electron-positron plasmas 
have been the subject of many papers over the last few decades.
Nevertheless, their role in pulsar winds is still in need of clarification. 
One of the sources of this confusion is rooted in the use of the \lq\lq strong
wave limit\rq\rq\ to simplify the treatment of linearly polarized waves
\citep{maxperkins71,kennelschmidtwilcox73,kennelpellat76}. This corresponds to 
taking only the lowest order terms in the parameter $q$ defined 
in (\ref{qdefinition}). To this order, the particle flux in the 
H-frame vanishes -- particles are said to be 
\lq\lq locked\rq\rq\ into the wave in this 
frame. Furthermore, the energy densities of particles and fields are
in equipartition to lowest order in $q$. These properties 
led \citet{kennelschmidtwilcox73}
and \citet{kennelpellat76} to assert that a strong wave would impose a 
rather severe upper limit on the particle flux from a pulsar.

The formulation we present in \S\ref{twofluid} and apply in \S\ref{results}
avoids this limit. It demonstrates that highly nonlinear
superluminal waves can carry a substantial particle flux, although the
minimum radius at which they can propagate moves outwards as the flux
increases.  These waves exhibit a wide range of $q$ values (see
Fig.~\ref{comparisonlincirc}). Particles are not locked into the wave
motion in the H-frame, and energy is not necessarily in equipartition.
Physically, the reason for this apparent contradiction is that 
$q$ does not adequately characterize the wave solution, since it 
is defined at a highly non-representative point in the 
phase of the wave; there exists a large family of 
nonlinear waves that have no counterpart with $q=0$.  

The question of the stability of superluminal waves has attracted
considerable attention. \citet{asseokennelpellat78} found that
radiation reaction would damp these waves within a few wavelengths, for
parameters appropriate to the Crab pulsar. However, this conclusion
was based on the $q=0$ limit discussed above, and may not apply to
large $q$ modes. Again working analytically in the limit $q=0$,
\citet{asseollobetschmidt80} found rapidly growing Weibel-like
instabilities, but noted that they are stabilized when the
phase-averaged magnetic field is sufficiently strong. For circular
polarization, \citet{leelerche78,leelerche79} also found rapidly
growing instabilities when the plasma is locked into the wave in the
H-frame. Particle-in-cell simulations have been used by several
authors to investigate this question. \citet{leboeufetal82} injected a
self-consistent wave and found it to be highly unstable when excited just
above the cut-off frequency, but much less so when excited at higher
frequencies. \citet{skjaeraasenmelatosspitkovsky05} injected a
nonlinear wave into a uniform plasma, and again found a strong
instability that could be partially stabilized in the presence of
streaming motion.  Although our results do not address this question
directly, it can be seen from the dispersion relations presented in
Figs.~\ref{circulardispersion} and \ref{lineardispersion} that the
H-frame moves relativistically with respect to the pulsar
($\Gamma_>\gg1$) for $R\gtrsim10$. An instability that
grows locally in the H-frame with growth-rate comparable to the wave
frequency in that frame, 
propagates a distance of $\Gamma_>^2$ wavelengths in the pulsar
reference frame in one $e$-folding time.  Thus, although such
instabilities may be present, their effect on the structure of the
pulsar wind or the termination shock might not be dramatic.

In \S\ref{results} we show that superluminal waves are capable of carrying 
the particle, energy and magnetic fields thought to be transported by 
a pulsar wind when they reach the termination shock of an isolated pulsar,
except for a range of latitudes around the (rotation) pole. In fact, at
most latitudes, there is a large range of radii between $\rho_{\rm cr}$ and 
$\rho_{\rm ts}$ in which propagation is possible. Our results do not 
address the question of where conversion from a striped wind to a
superluminal wave might occur. However, since these waves are subsequently 
damped, it would seem likely that they could be sustained only 
within an extinction length of the termination shock itself, in which case
their excitation and damping can be regarded as forming a part of this 
structure. 

\citet{lyubarsky03} suggested the physics of the \lq\lq termination
shock\rq\rq\ should differ from that of an MHD shock and permit
reconnection of the alternating component of the incoming magnetic
field.  A quantitative investigation based on analytic considerations
and 1D PIC simulations \citep{petrilyubarsky07} found that the
effectiveness of reconnection depends on the ratio of the Larmor radii
of electrons in the downstream flow to the wavelength of the
stripes. Since this quantity increases monotonically with radius in a striped
wind, these results imply that reconnection can be important only
outside of a critical radius.  Specifically, reconnection is predicted 
not to play a role for $R<\pi/12$ (in our notation). At larger
radii, the magnetic field is more effectively dissipated, until, at
$R>\pi\sqrt{\sigma}/3$, full dissipation is achieved. However, this
theory does not take account of the role of superluminal waves,
although these waves are able to propagate in the region
in which reconnection is predicted.

\citet{sironispitkovsky11}, on the other hand, performed 2D and 3D PIC
simulations. They found that the shock triggered reconnection, leading
to full dissipation of the magnetic field, over the entire parameter
range investigated. In our notation, this range extends over
$1/20\lesssim R\lesssim 4$. According to Fig.~\ref{lineardispersion},
superluminal waves of very high phase speed (small $\beta_>$) might be
expected to play a role at the upper end of this range. However, the
simulations do not appear to reveal their presence.  The particle
spectrum found in these simulations varies considerably, being closer
to that observed in the Crab Nebula for lower values of $R$.  However,
as we show in \S\ref{results}, the termination shocks in the nebulae
around isolated young pulsars are expected to lie much further from the
pulsar than the range covered by these simulations, leaving us free to
speculate that the physics controlling the structure and particle
acceleration process might be substantially different in these objects.

\section{Conclusions}

We have reexamined the properties of transverse superluminal waves in
an electron-positron plasma for parameters appropriate to pulsar
winds. Our principal new results are:

\begin{itemize} 
\item The properties of circularly and linearly polarized modes
  with vanishing phase-averaged magnetic field are similar
  (Fig.~\ref{comparisonlincirc}).
\item The widely used strong-wave approximation ($q=0$), which underlies
  the assertion that these waves limit the particle flux, is
  inadequate. There is no wave-intrinsic limit on the particle flux.
\item The cut-off surface, inside of which these waves cannot
  propagate, has a strong latitude dependence.
  (Fig.~\ref{boundaries}).  It lies well inside the termination shock
  for young isolated pulsars, but must cross it at some phase of the
  binary orbit in the case of PSR~B1259-63.
\item At high latitudes, where the phase-averaged magnetic flux
  carried by the wind is large, long wavelength, quasi-homogeneous
  modes of infinite phase speed (i.e., with $\beta>=0$,
  Fig.~\ref{lineardispersion}) can arise close to the cut-off surface.
\end{itemize}

On the basis of these results, and of recent discussions in the
literature of reconnection in the termination shock
\citep{petrilyubarsky07,sironispitkovsky11}, we speculate that
the physics of particle acceleration and the structure of this shock
may be strongly influenced by the presence of superluminal
waves.

\appendix
\section{Two-fluid equations}
\label{2fluideqns}

Following \citet{clemmow77} and \citet{kennelpellat76}, 
we present the solutions to the equations describing a cold, two-fluid 
plasma where {\it transverse} electromagnetic waves of superluminal phase 
speed can propagate. The equations are written in the H-frame and 
a plus (minus) index denotes positrons (electrons). It is 
convenient for the calculations to introduce complex quantities for the 
transverse four-velocity and the electric and magnetic fields:
\begin{eqnarray}
u_{\perp,\pm}=u_{y,\pm}+iu_{z,\pm} \\
E=E_y+iE_z \\
B=B_y+iB_z
\end{eqnarray}
Using these, the continuity equation and the equations of motion for each
species, expressed in the H-frame, are:
\begin{eqnarray}
\frac{d}{dt} \left( \gamma_{\pm} n_{\pm} \right) = 0 \label{continuity}\\
\gamma \frac{du_{\parallel,\pm}}{dt} = -\frac{e}{mc}{\rm Im}(u_{\perp,\pm} B^*) 
\label{motion1}
\\
\gamma \frac{du_{\perp,\pm}}{dt} = \frac{e}{mc}(\gamma E + i u_{\parallel,\pm}B) 
\label{motion2}
\\
\gamma \frac{d\gamma_{\pm}}{dt} = \frac{e}{mc} {\rm Re}(u_{\perp,\pm} E^*).
\end{eqnarray}where $e$ is the magnitude of the electron charge.

In the H-frame, all space derivatives
disappear. Conservation of magnetic flux, $\mathbf{\nabla}
\cdot \mathbf{B} = 0$, is
automatic, and the remaining Maxwell's equations (Ampere, Faraday and Coulomb
law, respectively) become:
\begin{eqnarray}
\frac{1}{c}\frac{dE}{dt} = - 4 \pi e (n_+ u_{\perp,+} - n_-u_{\perp,-}) \\
\frac{dB}{dt} = 0 \\
4\pi\rho = 4\pi e(n_+\gamma_+ - n_-\gamma_-) = 0. \label{coulomb}
\end{eqnarray}where $\rho$ is the charge density.
From (\ref{coulomb}), the charge density must vanish in the H-frame. In 
combination with the continuity equation, we get
\begin{eqnarray} 
n_+\gamma_+ = n_0\gamma_0 = n_- \gamma_- \enspace,
\nonumber
\end{eqnarray}where $n_0$,$\gamma_0$ are constants. The restriction to purely 
transverse electric fields (which is implied in the way Ampere's law is 
expressed above) means that the radial current must vanish: $n_+u_{x+} 
= u_{x-}n_-$. In this case, the net force acting on the plasma in the 
transverse direction vanishes, so that $u_{y,z+}+u_{y,z-}$ is constant. In
keeping with the approximation of toroidal fields (corresponding to the
$z$-direction in the plane wave approximation) and radial
propagation ($x$-direction), we set this constant to zero, so that $u_{y+}= - u_{y-}$
and $u_{z+}=-u_{z-}$. Because of these relations, we can restrict
ourselves to solving the equations for the positively charged fluid,
and will henceforth drop the plus subscript for simplicity.

\subsection{Circular polarization}

The simplest solution of these equations
is the case of a circularly polarized wave. Then $B=0$, and we have: 
\begin{eqnarray}
u_{\perp} = u_{\perp 0} {\rm e}^{i\omega_{\rm p}t} \\
E = E_{0} {\rm e}^{i\omega_{\rm p}t} \\
u_{\parallel} = u_0\\
n=n_0\enspace.
\end{eqnarray}
The frequency coincides with the (proper) plasma frequency: 
\begin{equation}
\omega_{\rm p}= \sqrt{\frac{8\pi n_0 e^2}{m}} 
\label{plasmafreqappendix}
\end{equation}
and the magnitude of the perpendicular component of the 
four-momentum equals the {\em strength} or 
{\em nonlinearity} parameter $\strength$, 
defined using the electric field amplitude as
\begin{eqnarray}
u_{\perp 0}=\strength\equiv\frac{e\left|E_0\right|}{mc\omega_{\rm p}}\enspace.
\end{eqnarray}
The parallel component of the four-velocity is a 
constant. 

In the laboratory frame,
which moves at speed $-c/v_{\phi}$ as 
seen from the H-frame, the frequency of the wave is 
\begin{equation}
\omega' =  \Gamma_>\omega_{p0} \label{freqlabframe}
\end{equation}
where $\Gamma_>$ is the Lorentz factor of the transformation between the H-frame and the lab.\ frame.

\subsection{Linear polarization}

Near the equatorial plane of a pulsar wind, the wave is linearly polarized 
and the relevant field components are
the polar electric field and the toroidal magnetic field. In the H-frame
then, this requirement corresponds to $E_z=0$
and $B_y=0$, i.e., $E=E_y$ becomes real and $B=iB_z$ purely imaginary (and
constant). It follows that the $z$-component of the four velocity is
constant $\textrm{Im}\left(u_\perp\right)$. In a pulsar, a non-zero $u_z$
corresponds to an azimuthal current loop, and, therefore, is
proportional to the integrated magnetic flux crossing the enclosed
surface. In the split-monopole geometry, this quantity falls off  
inversely with radius, and we set it to zero, so that $u_\perp$ is a 
real variable. 

Ampere's law becomes
\begin{eqnarray}
\frac{d E}{d \phi} = -\frac{mc\omega}{e}\frac{u_\perp}{\alpha \gamma}
\end{eqnarray}
where $\phi = \omega t$ and 
\begin{eqnarray}
\alpha = \frac{\omega^2}{\gamma_0 \omega^2_p}\enspace.
\end{eqnarray}
Multiplying the equations of motion (\ref{motion1}) and (\ref{motion2}) by 
$n$ and integrating, the four-velocity components can be 
expressed in terms of the variable $y=E/E_0$, the electric field normalized 
to the maximum value of its modulus (i.e., $-1\le y\le 1$) as follows:
\begin{eqnarray}
u_{\parallel} = u_0 + \alpha \strength^2 B_z (1-y)/E_0 \\
\gamma = \gamma_0 + \frac{\alpha \strength^2}{2}\left(1-y^2\right)\\
u^2_{\perp} = \gamma^2 - u^2_{\parallel} - 1\enspace. 
\label{norm4velocity}
\end{eqnarray}

Having expressed all other unknowns in terms of the electric field, it
remains to calculate the phase dependence of $y$, which is given by
the equation
\begin{eqnarray}
\alpha^2 \strength^2 \left(\frac{dy}{d\phi} \right)^2 
= \frac{u_{\perp}^2}{\gamma^2}
\end{eqnarray}
or, explicitly:
\begin{equation}
\alpha^2 \strength^2 \left( \frac{dy}{d\phi} \right)^2 
= \frac{N(y)}{[q/2 
+(1-y^2)]^2} 
\label{kennelequation}
\end{equation}
\citep{kennelpellat76}
with $q$, $Q$ and $b$ given by the expressions: 
\begin{eqnarray}
q = \frac{4 \gamma_0}{\alpha \strength^2} \label{justq}\\
Q = 2 \left(1 - \frac{u_0}{\gamma_0}  \lambda \right) \\
\lambda = \frac{B_z}{E_0}\enspace.
\end{eqnarray}
and 
\begin{eqnarray}
N(y)=(1-y)^2[(y+1)^2-4\lambda^2-q] +(1-y)qQ
\end{eqnarray}
is a quartic in y. Periodic waves exist if $N$ has four real roots,
which are (in order):
\begin{eqnarray}
y_1=\zeta\cos\left(\xi-4\pi/3\right)-1/3\\
y_2=\zeta\cos\left(\xi-2\pi/3\right)-1/3\\
y_3=1\\
y_4=\zeta\cos\left(\xi\right)-1/3
\end{eqnarray}
where 
\begin{eqnarray}
\zeta=2\left(4+12\lambda^2+3q\right)^{1/2}/3\\
\xi=\arccos\left[\frac{3\sqrt{3}}{\sqrt{2}\left(\sqrt{3} u\right)^{3/2}}
\left(16/27+2q/3-2 \lambda q u_0/\gamma_0-16\lambda^2/3\right)\right]
\end{eqnarray}
provided $\xi$ is real.

The nonlinear dispersion relation follows by demanding
that the change in phase in a half-cycle of the field oscillation
equal $\pi$:
\begin{equation}
\pi = \int_{y_1}^1 \frac{dy}{dy/d\phi}
\label{nonlineardispersion}
\end{equation}
The phase-averaged value of a quantity $A$ that depends on $y$ 
is given by 
\begin{equation}
\langle A \rangle = \frac{1}{\pi} \int_{y_1}^1 \frac{A(y)}{dy/d\phi}dy 
\label{means}
\end{equation}

If the phase-averaged magnetic field vanishes ($\lambda=0$), 
$N(y)$ reduces to a quadratic in $y^2$, with
$y_2=-1$ and $y_3=1$.  In this case,
the integrals in (\ref{nonlineardispersion}) and (\ref{means}) can be expressed
in closed form in terms of elliptic integrals, at least 
for the functions of interest.  
For non-zero $\lambda$, on the other hand,
$y$ oscillates between $y_2>-1$ and $y_3=1$. The dispersion relation and
the phase-averages must then be evaluated by numerical integration.

The components of the stress energy tensor consists of a fluid and a field
contribution. In the H-frame these are: 
\begin{eqnarray}
T^{00}_{\rm part} &=& 
2 m c^2 \langle n \gamma^2 \rangle 
\, =\, 2 n_0\gamma_0 mc^2 \langle \gamma \rangle 
\label{appT1}\\
T^{01}_{\rm part} &=& 
2 m c^2 \langle n \gamma u_x \rangle  
\,=\,  2 n_0\gamma_0 mc^2 \langle u_x \rangle 
\label{appT2} \\
T^{11}_{\rm part} &=& 
2 m c^3 \langle n u_x^2 \rangle 
\,=\,  2 n_0 \gamma_0 mc^3 \left< \frac{u_x^2}{\gamma}\right>
\label{appT3}
\end{eqnarray}
for the fluids, and
\begin{eqnarray}
T^{00}_{\rm EM} &=& 
\frac{\langle E^2 \rangle + B ^2}{8 \pi} 
\,=\, \frac{\langle y^2\rangle + \lambda^2}{8 \pi} E^2_0 \label{appTem1}\\
T^{01}_{\rm EM} &=& 
\frac{c\langle E \rangle B}{4\pi} 
\,=\, \frac{c\langle y \rangle \lambda E^2_0}{4\pi} \label{appTem2}\\
T^{11}_{\rm EM} &=& T^{00}_{\rm EM} \label{appTem3}
\end{eqnarray}
for the fields.
For the circularly polarized modes, 
$\lambda = 0$, $\left<y\right> = 0$, and $\left< y^2 \right> = 1$. In addition,
$u_x=u_{x0}$ and $\gamma=\gamma_0$ are phase-independent.

\section{Lorentz boosts}
\label{lorentzappendix}
The phase-averaged particle flux, energy, energy flux, and $x$-momentum flux densities and the electric field
transform according to:
\begin{eqnarray}
J'        &=& \Gamma_>  \left(2\beta_> n_0\gamma_0 +J \right) \\
{T'}^{00} &=& \Gamma_>^2 \left( T^{00} + \beta_> T^{01} + \beta_>^2 T^{11}\right)\\
{T'}^{01} &=& \Gamma_>^2 \left[ \left(1+\beta_>^2\right)T^{11} + \beta_> \left(T^{00}+T^{11}\right)\right] \\
{T'}^{11} &=& \Gamma^2_> \left(T^{11}+ 2\beta_>T^{01} + \beta_>^2T^{00}\right) \\
\left<{E'}\right>      &=& \Gamma_>  \left( \beta_>\lambda E_0 +  \left<E\right>\right)
\end{eqnarray}

\bibliographystyle{hapj}
\bibliography{references}
\end{document}